\documentclass[prl,showpacs,superscriptaddress,floatfix,twocolumn]{revtex4}
\usepackage[latin1]{inputenc}
\usepackage[english]{babel}
\usepackage{graphicx}
\usepackage{amsmath}
\usepackage{amssymb}
\usepackage{amscd}
\usepackage{eucal}
\usepackage{color}
\usepackage{bm}
\input {epsf}
\begin{document}

\title{Spin transistor action from Onsager reciprocity and SU(2) gauge theory}
\author{ \.{I}. Adagideli}
\affiliation{Faculty of Engineering and Natural Sciences, Sabanci University,
Orhanli-Tuzla, Istanbul, Turkey}
\author{ V. Lutsker}
\affiliation{Institut f\"ur Theoretische Physik, Universit\"at Regensburg, D-93040 Regensburg, Germany}
\author{ M.~Scheid}
\affiliation{Institut f\"ur Theoretische Physik, Universit\"at Regensburg, D-93040 Regensburg, Germany}
\author{Ph. Jacquod}
\affiliation{Physics Department and College of Optical Sciences, University of Arizona,
Tucson, AZ 85721, USA}
\affiliation{Theoretical Physics Department, University of Geneva, 1211 Geneva, Switzerland}
\author{K. Richter}
\affiliation{Institut f\"ur Theoretische Physik, Universit\"at Regensburg, D-93040 Regensburg, Germany}
\date{\today }

\begin{abstract}
We construct a local gauge transformation to show how, in confined systems,
a generic, weak nonhomogeneous
$SU(2)$ spin-orbit Hamiltonian reduces to two $U(1)$ Hamiltonians for spinless fermions
at opposite magnetic fields, to leading order in the spin-orbit strength. Using an Onsager
relation, we further show how the resulting spin conductance vanishes in a
two-terminal setup, and how it is turned on by either weakly breaking time-reversal symmetry or
opening additional transport terminals. We numerically check our theory
for mesoscopic cavities as well as Aharonov-Bohm rings.
\end{abstract}


\maketitle

Transistor action is often based on symmetries.
To switch on and off
a field effect transistor, an external gate turns a three-dimensional insulator into a two-dimensional metal and back. Compared to the off-state,
the on-state has thus reduced dimensionality and
symmetry. The relevance of symmetries in transistor action is even more
pronounced in some recently proposed spin-based transistors, whose action
follows directly from the breaking of spin rotational
symmetry, by tuning spin-orbit interaction (SOI) around a special symmetry point~\cite{sel},
where the SOI field reduces to two identical $U(1)$ fields with opposite coupling constants~\cite{duc10}.

In this manuscript, we propose a new class of spin
transistors whose action is
based on an Onsager reciprocity relation.
We show that in confined quantum coherent systems with spatially inhomogeneous
SOI (Rashba, Dresselhaus or impurity SOI, or a combination of the three),
an appropriate $SU(2)$ gauge transformation allows to express the
spin conductance $G_{ij}^\uparrow-G_{ij}^\downarrow$ between two terminals labeled
$i$ and $j$ through the charge magnetoconductance $ G_{ij}({\cal B})$ as
$G_{ij}^\uparrow-G_{ij}^\downarrow
= G_{ij}({\cal B}) - G_{ij}(-{\cal B})$. This holds to leading order in the
ratio $L/\ell_{\rm so} \ll 1$ of the system
size $L$ and the spin-orbit (precession) length $\ell_{\rm so}$. The
pseudo magnetic field ${\cal B}$ arises from the gauge-transformed SOI.
Current conservation together with
the Onsager relation $G_{ij}({\cal B})=G_{ji}(-{\cal B})$~\cite{Onsager,Buttiker86}
then forces $G_{ij}^\uparrow-G_{ij}^\downarrow=0$ to leading order
for a two-terminal setup.
This is the off state of our transistor. The on state is obtained by either opening
additional terminals
or breaking time-reversal symmetry with a true
magnetic field $B_0$, in which case
$G_{ij}^\uparrow-G_{ij}^\downarrow = G_{ij}(B_0+{\cal B}) - G_{ij}(B_0-{\cal B}) \ne 0$,
even in a two-terminal setup.
Our {\it Onsager spin transistor} can thus be controlled either electrically or magnetically.
In both instances, this turns on
a spin conductance $G_{ij}^\uparrow-G_{ij}^\downarrow  \propto \ell_{\rm so}^{-1}$ with
an on/off ratio $\propto (\ell_{\rm so}/L)^2 \gg 1$.
The mechanism works in
diffusive as well as ballistic systems, and is more pronounced in regular systems with
few channels.

Aleiner and Falko constructed a
gauge transformation to show that, in confined systems with $L/\ell_{\rm so} \ll 1$,
a homogeneous $k$-linear SOI has a much weaker effect on
charge transport than the naive expectation $\propto \ell_{\rm so}^{-1}$~\cite{Aleiner}.
Brouwer and collaborators later argued
that
terms $\propto \ell_{\rm so}^{-1}$ in the charge conductance
survive the gauge transformation for
SOI with spatially varying strength~\cite{Brouwer}.
The relevance of the  pseudo magnetic field for a specific mesoscopic
system with inhomogenous SOI was noticed in Ref.~\cite{SU2}.
It is however not clear how much of the gauge arguments of Refs.~\cite{Aleiner,Brouwer,Tokatly}
carry over to spin transport in generic systems~\cite{Ada10}. Below we show
that gauge transformations result in different symmetries for charge and for
spin transport~\cite{comment-charge-cond}.

Our starting point is a two-dimensional Hamiltonian for electrons with
SOI, which we write as ($\hbar \equiv 1$)
\begin{equation}\label{Hamsu2}
H=-\frac{1}{2m}D_{\mu }D_{\mu }+V(\mathbf{x}) \; .
\end{equation}%
Here, $V({\bf x})$ is a spin-diagonal potential and
the covariant derivative $D_{\mu }=\partial _{\mu }-(i k_{\rm so}/2)\sigma^{a}A_{\mu }^{a}$
contains the SOI via the $SU(2)$ gauge field
$\sigma^a
A_{\mu }^{a}$, with the Pauli matrix $\sigma^a$.
From here on,
Latin indices are spin indices, while Greek letters denote
spatial indices. The
SOI
constant $k_{\rm so}$ determines the spin-orbit length as
$\ell_{\rm so}=\pi |k_{\rm so}|^{-1}$. We consider a gauge transformation
${\cal O}\rightarrow {\cal O}'=U {\cal O} U^{-1}$
with $U=\exp(i{\sigma^a}{\Lambda}^a/2) \simeq 1+i \sigma^a \Lambda^a/2$, and
search for a $\Lambda^a$ that
reduces the leading order, $k_{\rm so}$-linear part of the SOI to a
spin-diagonal $U(1) \times \sigma^z$ structure. We
use the well-known decomposition ($\epsilon_{\mu \nu}$ is the totally antisymmetric
tensor of order two) for each spin component
\begin{equation}
A_{\mu }^{a}=-(\partial _{\mu }\chi ^{a}+\epsilon _{\mu \nu }\partial _{\nu
}\phi ^{a}) \; ,
\end{equation}%
with $\phi^a$ given by $\nabla^{2}\phi ^{a}=\epsilon _{\mu \nu }\partial _{\mu }A_{\nu }^{a}$.
In particular, $\phi ^{a}$ is necessarily nonzero for spatially varying SOI. It is
straightforward to see that the choice $\Lambda^a = k_{\rm so} \chi^a$ gauges away
the gradient part of the $SU(2)$ vector potential to linear order in
$k_{\rm so}$,
\begin{eqnarray}
\label{gaugefield}
A_{\mu }^{a} & \rightarrow & (A^{\prime })_{\mu }^{a}
=-\epsilon _{\mu \nu }\partial _{\nu }\phi ^{a}+\mathcal{O}(k_{\rm so}) \; .
\end{eqnarray}%
Note that $\mathcal{O}(k_{\rm so})$ corrections in $A_{\mu}^a$ lead to
$\mathcal{O}(k_{\rm so}^2)$ corrections in the Hamiltonian.
If the SOI strength is spatially constant, $\phi^a=0$ and one recovers
the result of Ref.~\cite{Aleiner} that all ${\cal O}(k_{\rm so})$-terms are gauged away.

We next want to extract the leading order, linear
in $k_{\rm so} \chi^a \ll 1$
dependence of transport properties such as conductances, and thus use
\begin{equation}
{\cal O}'=U {\cal O} U^{-1}={\cal O}+ik_{\rm so}[{\sigma^a}\chi ^{a},{\cal O}]/2 \; .
\end{equation}
In particular we have $D_{\mu }\rightarrow D_{\mu }^{\prime }$ with
\begin{equation}\label{spinrot}
\sigma^a \rightarrow {\sigma^a}'=\sigma^a+ k_{\rm so}
\epsilon^{abc} \chi ^{b}({\bf x}) \sigma^c \, .
\end{equation}
To calculate spin conductances we need to gauge transform the operator for spin current
through a cross-section $C_j$ in terminal $j$,
$\hat{I}^a_j=\int_{C_j} d\alpha \, \{{\bf n}_\alpha \cdot{\bf j}(\alpha),\sigma^a\}$.
We obtain
\begin{eqnarray}\label{spinop}
(\hat{I}^\prime)^a_j
&=&\int_{C_j} d\alpha \, \left[ \{{\bf n}_\alpha \cdot{\bf j}'(\alpha),\sigma^a +
k_{\rm so} \, \epsilon^{abc} \chi^{b}(\alpha) \sigma^c\} \right] \nonumber \\
&=& \hat{J}^a_j+k_{\rm so} \delta \hat{J}^a_j
\end{eqnarray}
where $\hat{J}^a_j$ is the ``naive'' spin current of the transformed Hamiltonian, not accounting for the
rotation (\ref{spinrot}) of the spin axes.
We further need the Heisenberg picture operators $\hat{I}^a_j(t)={\rm e}^{i H t} \hat{I}_j^a{\rm e}^{-i H t}$ which
transform as
\begin{eqnarray}
(\hat{I}^\prime)^a_j(t)
&=& \hat{J}^a_j(t) +k_{\rm so} \delta \hat{J}^a_{j0}(t)+{\cal O}(k_{\rm so}^2) \, .
\label{EQ:gauge_trans_SC}
\end{eqnarray}
Here the subscript $0$ means that the time-evolution is through the $k_{\rm so}=0$ Hamiltonian.

Linear response relates chemical potentials in external reservoirs and currents
in the leads via the spin-conductance
matrix as $I_{i}^a=\sum_{j}G_{ij}^{a}\mu_{j}/e$.
It is somehow tedious, though straightforward to
show that, to linear order in $k_{\rm so}$,
the gauge transformation gives $G_{ij}^a \rightarrow (G^\prime_{ij})^a$, with
the conductance matrix $(G^\prime_{ij})^a$ evaluated in the same way as $G_{ij}^a$ but
with the spin current operators $\hat{J}^a_j$ of the transformed Hamiltonian in Eq.~(\ref{EQ:gauge_trans_SC}).
Thus, to leading order in $k_{\rm so}$, infinitesimal nonabelian gauge
transformations preserve the form of the spin conductance.
Note that global gauge transformations (i.e. global spin rotations),
whether infinitesimal or finite,
are easy to introduce via the corresponding rotation matrix ${\cal R}$ as
$G_{ij}^a =  {\cal R}^{ab} (G_{ij}')^b$. All global or local spin gauge transformations leave
the potential $V({\bf x})$ invariant.

We are now equipped to use the gauge transformation to explore the spin conductance.
We first focus on the exactly solvable case of a Rashba SOI~\cite{rashba}
with a spatially varying strength $\alpha({\bf x}) = k_{\rm so} \,
\overline{\alpha} \, ({\bf x}\cdot\mathbf{f} )$, with a dimensionless function
$\overline{\alpha }$, whose gradient always points in the direction of the unit vector
$\mathbf{f}$.
One has $A_{\mu }^{a} =-2 \overline{\alpha }(\mathbf{x}\cdot\mathbf{f})\epsilon _{a\mu }$,
$\phi^{a}(\mathbf{x})=\varphi (\mathbf{x})f^{a}$, and Eq.~(\ref{gaugefield}) gives
\begin{subequations}
\begin{eqnarray}
(A^{\prime })_{\mu }^{a} &=&-\epsilon _{\mu \nu }\partial _{\nu }\varphi (%
\mathbf{x})f^{a}+\mathcal{O}(k_{\rm so}) \; , \\
D_{\mu }^{\prime } &=&\partial _{\mu }+\frac{i}{2}k_{\rm so} \epsilon _{\mu \nu }\partial
_{\nu }\varphi (\mathbf{x}){\boldsymbol{\sigma \cdot }}\mathbf{f.}
\end{eqnarray}%
\end{subequations}
After the global spin rotation
${\boldsymbol{\sigma\cdot }}\mathbf{f\rightarrow } \sigma^{z}$,
Eq.~(\ref{Hamsu2})
becomes
\begin{subequations}
\begin{eqnarray}
H&=&\left(
\begin{array}{cc}
h(\boldsymbol{a}) & 0 \\
0 & h(-\boldsymbol{a})%
\end{array}%
\right)   + {\cal O}(k_{\rm so}^2) \label{TransHam} \; , \\
h(\boldsymbol{a})&=&-\frac{1}{2m}\left[
\nabla +ik_{\rm so}\boldsymbol{a}(\mathbf{x})
\right]^2
+V(\mathbf{x}) \; .
\end{eqnarray}%
\end{subequations}
Thus to linear order in $k_{\rm so}$, the Hamiltonian is mapped onto
a block spin Hamiltonian where the opposite spins feel opposite, purely orbital pseudo
magnetic fields
${\mathcal{B}} =(\nabla\times\boldsymbol{a})_z
$ generated by the $U(1)$
vector potential
$a_{\mu }=\frac{1}{2}\epsilon _{\mu \nu }\partial _{\nu}\varphi (\mathbf{x})$.
We obtain
$\mathcal{B}(\mathbf{x})=k_{\rm so} \mathbf{f}\cdot\nabla \bar{\alpha}$.
Transforming back to the original gauge, the spin
conductance
is obtained as
$
G_{ij}^a=\big[ G_{ij}(\mathcal{B})-G_{ij}(-\mathcal{B}) \big] f^a + {\cal O}(k_{\rm so}^2).
$
In this simple example, the spin conductance is thus the difference
of two charge conductances $ G_{ij}$ at opposite pseudo magnetic fields.
For generally varying SOI,
one cannot choose a spin quantization axis as before. Thus we need to define one
pseudo-magnetic field per spin polarization, i.e.~we define
${\cal B}_{a}=\partial_x A_y^a-\partial_y A_x^a$ as the magnitude of a pseudo
magnetic field (pointing always in $z$-direction) that arises solely from the $a$ component
of $\phi ^{a}$. To linear
order in $k_{\rm so} L$, the superposition principle gives the spin conductance
along axis $a$ as solely due to the
component of $\phi ^{a}$, $G_{ij}^a= G_{ij}({\cal B}_{a})-G_{ij}(-{\cal B}_{a})$.
The same argument gives
the leading-order spin conductance in the presence of an externally applied
(i.e. true) magnetic field $B_0$ as
\begin{equation}
G_{ij}^a(B_0)= G_{ij}(B_0+{\cal B}_{a})-G_{ij}(B_0-{\cal B}_{a})+{\cal O}(k_{\rm so}^2).
\label{spin-charge-conductance}
\end{equation}
This is our main result. It expresses the
{\sl spin} conductance of the original dot with SOI in terms of {\sl charge} conductances of the dot without
SOI, but with effective magnetic fields $B_0 \pm {\cal B}_a$ arising from the true applied
field, $B_0$, and
the pseudo field, ${\cal B}_{a}$, generated by the gauge transformation and
the SOI.

The key observation is then that the reciprocity relation
$G_{ij}(B) = G_{ji}(-B)$~\cite{Buttiker86},
together with gauge invariance,
$\sum_j G_{ij}(B) = 0$, imply that
the spin conductance (\ref{spin-charge-conductance}) vanishes to order
$\mathcal{O}(k_{\rm so})$ in two-terminal geometries
in the absence of external magnetic field,
since only then $G_{ij}(-{\cal B}_a)\!=\! G_{ji}({\cal B}_a)\!=\! G_{ij}({\cal B}_a)$.
On the contrary, $G_{ij}^a$ is linear in $k_{\rm so}$, i.e.
much larger,
when an external magnetic field is applied or when one (or more) additional
terminals are open.
Thus, multi-terminal spin conductances
linearly depend on $k_{\rm so} L$,
whereas two-terminal local conductances are quadratic or higher order in
$k_{\rm so} L$. These restrictions imply that any coherent conductor
with spatially varying SOI can be operated as a spin transistor, whose action
is controlled by either opening an extra terminal or applying an
external magnetic field. This is the fundamental
mechanism on which the Onsager spin transistor we propose is based.

\begin{figure}
 \centering
 \includegraphics[width=0.95\columnwidth]{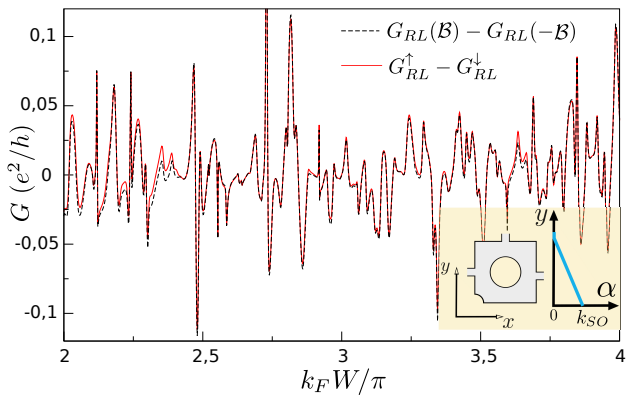}
\caption{
Comparison of the spin conductance $G_{RL}^y \!=\!G^\uparrow_{RL} \!-\! G^\downarrow_{RL}$
with the difference in the magnetoconductance, Eq.~(\ref{spin-charge-conductance}),
for transport (from left to right lead) through the three-terminal ballistic
quantum dot (see inset) with linear size $L$, leads of width $W$
and spatially varying SOI $\alpha({\bf x}) \!=\!k_{\rm so} \bar{\alpha}({\bf x}) =
k_{\rm so}(y/L) $ ({\em i.e.} ${\cal B} \!=\! \partial_y \alpha \!=\! k_{\rm so} / L$)
with $k_{\rm so} L \approx 0.3$.
}
\label{Fig:gauge-argument}
\end{figure}
%

\begin{figure}
 \centering
 \includegraphics[width=1.0\columnwidth]{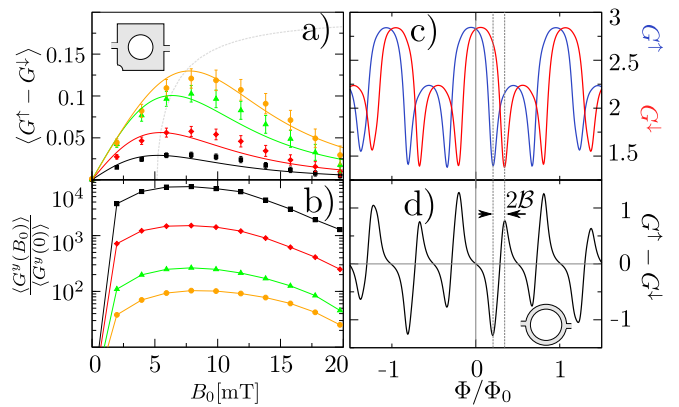}
\caption{
Spin conductances for two-terminal geometries as a function of
an applied magnetic field $B_0$.
(a) Average spin conductance of a chaotic cavity (inset)
for four different strengths of a linearly varying SOI (same as inset
Fig.~\ref{Fig:gauge-argument}) with
$ k_{\rm so}  L \approx 0.16, 0.33, 0.67$ and 1.0 from bottom to top curve.
Symbols with statistical errorbars mark numerical results for the average spin conductance,
full lines depict the theoretical prediction (\ref{weak-loc}).  The
grey dashed line shows predicted spin conductance maxima (from Eq.~(\ref{weak-loc})) for
varying gauge field.
(b) corresponding on-to-off ratios $\left\langle G^y(B_0)\right\rangle / \left\langle G^y(0) \right\rangle$.
(c) spin resolved conductances $G^{\uparrow(\downarrow)}(\Phi)$
for an AB ring (inset panel (d))
as a function of flux $\Phi \!=\! \pi R^2 B_0$, showing a shift $\pm {\cal B}$ due to the
gauge field ${\cal B}=\nabla \alpha$
arising  from SOI $\alpha \!=\! (y/L)  k_{\rm so}$ with $ k_{\rm so} L \!=\! 1$.
(d) resulting spin conductance $ G^y(\Phi)$ of a single AB ring.
Inset panel (a): Sinai-type billard: linear size $L$, stopper disk with radius $R_i = L/10$, leads of width $W = L/15$ hosting
4 transverse channels.
Inset panel (d): AB ring: radius $R=L/2$, width $W = L/15$ with 4 open channels.
}
\label{Fig:AB}
\end{figure}

\begin{figure}
 \centering
 \includegraphics[width=1.\columnwidth]{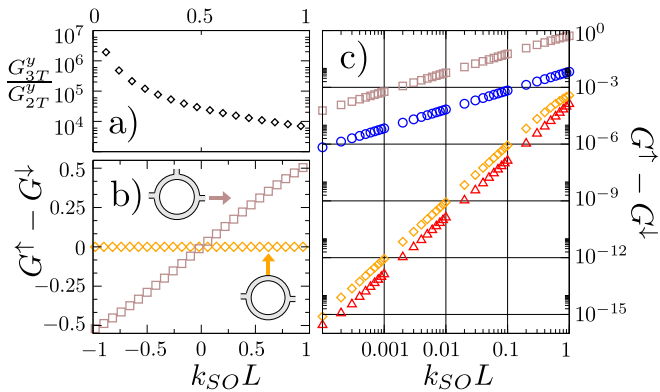}
\caption{
Onsager symmetry-based transistor action resulting from the difference in spin conductance
of two- and three-terminal mesoscopic rings (insets panel (b)).
(a) On--off ratio and (b) separate spin conductances  $G^y_{\rm 3T}$ and $G^y_{\rm 2T}$
for AB ring in three- and two-terminal mode as a function of a spatially nonuniform SOI,
$\alpha= (y/L)  k_{\rm so} $.
(c) double-log plot of same data as in (b) (top and third symbol sequence)
and of corresponding  $G^y_{\rm 3T}$ and $G^y_{\rm 2T}$ (second and fourth sequence) for
a more generic nonuniform SOI $\alpha =  k_{\rm so} \cos^2(2\pi x/L_1) \cos^2(2\pi y/L_2)$.
}
\label{Fig:Onsager-transistor}
\end{figure}

We numerically confirm these results by computing~\cite{Wimmer09}
the charge and spin conductances for two- and three-terminal
mesoscopic cavities and rings (sketched in the inset of
Figs.~\ref{Fig:gauge-argument}--\ref{Fig:Onsager-transistor}).
We first assume a Rashba SOI with constant gradient over the whole conductor,
$\alpha({\bf x}) \!=\! (y/L) k_{\rm so}$, and check
the prediction~(\ref{spin-charge-conductance}) that
the spin conductance can be expressed in terms of the charge conductance of the
transformed system without SOI but with a magnetic field ${\cal B}$.
In Fig.~\ref{Fig:gauge-argument},
the spin conductance  $G^y_{RL} \!=\! G^\uparrow_{RL} - G^\downarrow_{RL}$
(from now on the $y$-axis is the spin quantization axis) in the
absence of magnetic field is
compared to the difference of the charge conductance,  $G_{RL}({\cal B}) \!-\! G_{RL}(-{\cal B})$
in the absence of SOI, but with magnetic field ${\cal B}= \partial_y \alpha$.
Both quantities exhibit precisely the same mesoscopic conductance fluctuations as a function
of Fermi momentum, as predicted by Eq.~(\ref{spin-charge-conductance}).
We found that this level of agreement holds up  to $k_{\rm so} L  \approx 1 $,
beyond which terms quadratic and higher order in $k_{\rm so}$ are no longer subdominant.

%

For weak magnetic fields
(with an  associated cyclotron radius larger
than $L$), $G_{ij}(B)-G_{ij}(-B)$ is
predominantly given by quantum coherent contributions only. They give rise, on top of
the mesoscopic fluctuations displayed in Fig.~\ref{Fig:gauge-argument}, to a shift
$\delta G$ in the (energy) averaged conductance, known as weak localization correction.
In the presence of a magnetic field, $\delta G$ exhibits a damping that is Lorentzian-like,
$ \delta G (B) = \delta G(0) /(1+\xi B^2) $,
for chaotic ballistic cavities~\cite{Bar93}
with $\delta G(0) \sim (1/4) e^2 / h$ and $\xi$ proportional to the dwell time in the
cavity. According to  the prediction (\ref{spin-charge-conductance}) for the two-terminal
case, the presence of an external magnetic field $B_0$ leads to
a finite spin conductance $G^y(B_0) = G(B_0+{\cal B})-G(B_0-{\cal B})$, with
${\cal B} =  \partial_y\alpha$. Then its energy average is
\begin{equation}
\langle G^y(B_0) \rangle = \frac{\delta G(0)}{1+\xi(B_0+{\cal B})^2} -
\frac{\delta G(0)}{1+\xi(B_0-{\cal B})^2}
\, .
\label{weak-loc}
\end{equation}
This line of reasoning is confirmed in Fig.~\ref{Fig:AB}(a)
where numerically calculated spin conductances (symbols) for the chaotic cavity with linearly
varying SOI are compared to the prediction (\ref{weak-loc}) (full lines).
Figure ~\ref{Fig:AB}(b) shows the corresponding on-off ratios
$\langle G^y(B_0) \rangle / \langle G^y(0) \rangle $.

Alternatively, we consider
few-channel regular Aharonov-Bohm (AB) rings where $k_{\rm so}$-linear
spin currents can be turned on by a magnetic flux~\cite{SU2}. These systems
exhibit large almost periodic AB conductance oscillations instead of the weaker,
randomly-looking conductance fluctuations.
In Fig.~\ref{Fig:AB}(c) we present numerically computed spin resolved conductances
$G^{\uparrow(\downarrow)}(\Phi)$ as a function of flux $\Phi = \pi R^2 B_0$
(in units of the flux quantum
$\Phi_0 =h/e$) for an AB ring (inset panel (d)) in presence of the same linearly varying
SOI as for the cavity. As expected, the conductance traces for the spin-up and -down
channels are shifted against each other by
$\pm {\cal B} \!=\! \pm\partial_y\alpha$. This shift gives rise to a finite $B_0$-periodic
spin conductance $G^y \!=\! G^{\uparrow}\!-\!G^{\downarrow}$
as displayed in Fig.~\ref{Fig:AB}(d). At $B_0\!=\!0$,
first order spin conductance is forbidden by the Onsager relation.
 $G^y$ vanishes further for fields corresponding to $\Phi_0, \Phi_0/2$ and $\Phi_0/4$,
where maxima and minima of the usual charge magnetoconductance occur.
Maxima of $G^y$ appear at points where the shifted spin resolved
$G^{\uparrow(\downarrow)}$ have their minima.
This holds
for regular, or quasi-regular electronic dynamics
which requires clean AB rings with few-channels.
Of particular interest in the AB case are: (i) the
magnitude of the spin conductance, which exceeds its value in chaotic systems by one
to two orders of magnitude (compare the vertical axes scales in
Fig.~\ref{Fig:gauge-argument} and \ref{Fig:AB}(d)), and (ii) the control one has over the
spin conductance: Applying an integer or half-integer
flux quantum gives the off state of our transistor, while the on state is recovered
at $ B_0\!=\!  \pm[(\Phi_0/4)/ (\pi R^2) \!-\!  {\cal B}] $.
The on/off spin current ratio can be made arbitrarily large,
as it exactly vanishes in the off state.

As said above, $k_{\rm so}$-linear spin conductances can also be turned
on by adding an additional terminal.
As shown in Fig.~\ref{Fig:Onsager-transistor}(a,b) we find a  difference of at least
three orders of magnitude in spin conductance, $G^y_{\rm 2T}$ vs.\ $G^y_{\rm 3T}$,
for two- and three-terminal rings.
In panel (c) a double log representation of the data from (b) reveal the cubic
vs.\ linear $k_{\rm so} L$ dependence of $G^y_{\rm 2T}$ (top symbol sequence in
(c)) and $G^y_{\rm 3T}$ (third sequence from top) in line with our predictions.

So far we have considered linearly varying SOI. However, our theory holds generally
and works well also for more generic spatial dependence of the SOI.
We confirm this by calculating $G^y ( k_{\rm so} L)$  for a ring with SOI
$\alpha(\mathbf{r}) =  k_{\rm so} \cos^2(2\pi x/L_1)\cos^2(2\pi y/L_2)$ with
$L/L_1 \!=\! 15, L/L_2 \! =\! 6$ giving rise to SOI bumps on scales of the ring width.
As demonstrated in  Fig.~\ref{Fig:Onsager-transistor}(c)
we recover again the linear vs.\ cubic scaling with $ k_{\rm so} L$ for the
two- and three-terminal setting (second and fourth symbol sequence from top), in full
accordance with our theory.

%

We conclude with a few remarks:

(i)
Mesoscopic rings based on InAs~\cite{Berg06} or p-doped GaAs samples which are known
to exhibit large and tunable SOI~\cite{ensslin} are excellent candidates to
experimentally probe our theory. Inhomogeneous SOI could, e.g., be realized
through a  top gate covering only part of the system.
Additionally, a measurement protocol for spin currents
based on  symmetries of charge transport through quantum point contacts~\cite{Stano}
could be implemented.

(ii)
Inhomogeneous SOI is also a prerequisite for various specific proposals for spin
splitting~\cite{Kho04,Sun05}
and analogues of the Stern-Gerlach effect~\cite{Ohe05}. Our theory provides a rather general,
common footing to interpret them. For instance, the Stern-Gerlach based spin separation,
usually explained in terms of a Zeeman coupling in a non-uniform (in-plane) magnetic field (associated
with Rashba SOI), finds its explanation in the opposite bending of electron paths owing to the
Lorentz force associated with our gauge field $\pm {\cal B}$.

(iii)
Another gauge transformation, dual to ours, allows to transform a  nonuniform
Zeeman term into two decoupled components with an additional gauge field~\cite{Kor77}.

(iv)
While the spin conductance fluctuations are similar in a (phase coherent) diffusive system,
its classical magnetoconductance has a linear in magnetic field
contribution originating from the classical Hall effect. Thus  in a diffusive system with
inhomogeneous SOI, we expect a spin conductance with a nonzero average value
proportional to the classical Hall conductance.
This spin conductance can be estimated~\cite{Ada-et-al}
 as $\langle G^a\rangle  \sim (e^2/h) (k_{\rm so} \ell)$ where $\ell$ is the mean free path.
We stress that $\langle G^a\rangle $ is based on a classical effect in that it is robust against
effects such as dephasing and temperature broadening.

We thank M.~Duckheim for carefully reading our manuscript, and D.~Loss, J.~Nitta and M.~Wimmer for helpful conversations.
This work was supported by TUBITAK under grant 110T841 and the funds of the Erdal \.{I}n\"{o}n\"{u} chair (IA),
by NSF under grant DMR-0706319 and the Swiss Center for Excellence MANEP
(PJ), and by DFG within SFB 689 (MS,KR).


\begin{thebibliography}{9}
\bibitem{comment-charge-cond}
Our gauge transformation gives $G({\cal B})+G({-\cal B})$ for the charge conductance,
which is a linear function
of $\ell_{\rm so}^{-1}$, in agreement with Ref.~\cite{Brouwer}, regardless of the geometry.
\end{thebibliography}
\end{document}